# ARTICLE

## Label-free Prediction of Vascular Connectivity in Perfused Microvascular Networks in vitro

Liang Xu,[1ab] Pengwu Song,[1ab] Shilu Zhu,[ab] Yang Zhang,[c] Ru Zhang,[ab] Zhiyuan Zheng,[c] Qingdong Zhang,[c] Jie Gao,[ab] Chen Han,[ab] Mingzhai Sun,[ab] Peng Yao,*[d] Min Ye*[ab] and Ronald X. Xu*[abc]

Continuous monitoring and in-situ assessment of microvascular connectivity have significant implications for culturing vascularized organoids and optimizing the therapeutic strategies. However, commonly used methods for vascular connectivity assessment heavily rely on fluorescent labels that may either raise biocompatibility concerns or interrupt the normal cell growth process. To address this issue, a Vessel Connectivity Network (VC-Net) was developed for label-free assessment of vascular connectivity. To validate the VC-Net, microvascular networks (MVNs) were cultured in vitro and their microscopic images were acquired at different culturing conditions as a training dataset. The VC-Net employs a Vessel Queue Contrastive Learning (VQCL) method and a class imbalance algorithm to address the issues of limited sample size, indistinctive class features and imbalanced class distribution in the dataset. The VC-Net successfully evaluated the vascular connectivity with no significant deviation from that by fluorescence imaging. In addition, the proposed VC-Net successfully differentiated the connectivity characteristics between normal and tumor-related MVNs. In comparison with those cultured in the regular microenvironment, the averaged connectivity of MVNs cultured in the tumor-related microenvironment decreased by 30.8%, whereas the non-connected area increased by 37.3%. This study provides a new avenue for label-free and continuous assessment of organoid or tumor vascularization in vitro.

## Introduction

The vascular system in the human body is responsible for the transport of nutrients, oxygen, and metabolic waste products. When the integrity of the body's vasculature is compromised, resulting in impaired connectivity, it can be a limiting factor in the study and treatment of diseases such as atherosclerosis, coronary heart disease, myocardial infarction and cancer.[1-3] For instance, atherosclerosis can result in the formation of plaques that narrow or obstruct the lumen, thereby impairing the normal flow of blood and reducing the proper connectivity between blood vessels.[4, 5] Coronary artery disease and myocardial infarction can result in ischemia of the entire body or localized tissues. This results in impaired vascular connectivity and blood flow, which ultimately leads to damage to tissue function.[6, 7] In recent years, there has been a widespread demand in the field of tissue engineering to cultivate larger and more mature organoids and tumors. In this context, the vascular system is of paramount importance.[8-10] The research has demonstrated that in the absence of a perfusable vascular system, organoids and tumors are reliant on passive diffusion through the culture medium to transport of substances. However, the effective diffusion distance is limited to 300 μm, which gradually leads to the formation of necrotic cores and ultimately causes the organoid to fail to grow.[11, 12] Consequently, efforts have been made to establish connections between blood vessels and organoids or tumors with the objective of ensuring the supply of nutrients and the elimination of metabolic waste.[13-15] The study of vascular connectivity in vascularized organoids or tumors is crucial to understand the role of blood vessels in facilitating the exchange of nutrients and metabolic waste.

The complexity of the in vivo vascular system and the difficulty in studying it, coupled with the limitations of animal models in accurately reflecting human physiological responses due to species differences, have led to an increase in the popularity of research on in vitro cultured vascular models.[16] Initially, researchers constructed in vitro vascular models based on two-dimensional (2D) models of ECs. One limitation of these models is their inability to replicate the three-dimensional (3D) structure and the complex microenvironment of interactions between different cell types, which are crucial for vascular research.[17, 18] In recent years, in vitro microvascular networks (MVNs) have been formed by suspending ECs and other stromal cells in hydrogels within microfluidic device. These networks have been used to simulate ECs migration, blood-brain barrier formation, organoid development, and tumor vascularization.[13, 19, 20] This approach allows researchers to observe and study the human vascular system under conditions that more closely

[a.] School of Biomedical Engineering, Division of Life Sciences and Medicine, University of Science and Technology of China, Hefei, 230026, China.
[b.] Suzhou Institute for Advanced Research, University of Science and Technology of China, Suzhou 215123, China.
[c.] Department of Precision Machinery and Instrumentation, School of Engineering Science, University of Science and Technology of China, Hefei 230027, China.
[d.] School of Microelectronics, University of Science and Technology of China, Hefei 230027, China.
[1.] Co-first authors
[*.] Corresponding Authors

resemble those in vivo. However, these MVNs structures are often disorganized, with some parts being interconnected and others not, making it challenging to visually discern the connectivity of the vessels.[21] In previous studies, the method employed to determine the connectivity of vessels within MVNs involved perfusing fluorescent dextran dyes[13, 16, 22-24] or fluorescent microspheres[8, 24-26] into the vascular models. Subsequently, the distribution of the fluorescent dyes or microspheres within the vessels was observed through microscopic imaging in order to assess the connectivity between the vascular networks. Nevertheless, prior research has demonstrated that the incorporation of fluorescent dextran and fluorescent microspheres can result in an imbalance of intracellular inflammatory or oxidative stress markers, thereby introducing ambiguity to the investigation of vascularized organoids or tumors during continuous culture.[27-29] Furthermore, fluorescent perfusate can also remain in vascularized models such as vascularized organoids or tumors. When specific fluorescent markers in the vascularization models are to be observed in real time, the light emitted by the fluorescent perfusion fluid may be erroneously detected in the channel of the fluorescent marker, which may result in optical crosstalk and lead to inaccurate observation and detection results.[30, 31] Moreover, if a single perfusion is followed by the restarting of the culture of these in vitro vascularized models, both the time cost, which can span several weeks, and the economic cost are considerable. Additionally, in this case, the reproducibility of the experiment such as organoid culture is low, and new experiments cannot fully replicate the growth conditions of the cells from the previous experiment.[32, 33] Consequently, the accurate and low-cost prediction of MVNs connectivity, without the utilization of fluorescent dyes and without compromising vascularization models, will facilitate more efficient studies of organoid or tumor development.

In recent years, deep learning techniques have been employed in the study of in vitro cultured MVNs. To accurately quantify the 3D morphology of vessels, Ju et al.[34] employed an enhanced graph convolutional network, while Choi et al.[35] integrated U-Net and other network architectures with conventional image algorithms to segment the MVNs skeleton and ECs cultured on Matrigel.[36] Additionally, Kim et al.[37] developed Angio-Net, based on a generative adversarial network, which enables the reconstruction of fluorescent images of vascular sprouting from label-free bright-field images. This eliminates the need for complex immunofluorescence staining procedures. Mathur et al.[38] employed finite element methods and segmentation algorithms to predict the spatial distribution of oxygen and other physiologically relevant transport metrics within MVNs and surrounding tissues. Despite the aforementioned advancements, the widespread integration of deep learning technology in the field of in vitro cultured vessels remains limited. Currently, there is a paucity of research techniques for the label-free prediction of the connectivity of in vitro vascular models with perfusion functionality. Consequently, the urgent issue of how to train an accurate connectivity prediction model with low-cost, requiring only a small amount of connectivity data samples, must be addressed.

In this study, we established a deep learning-based label-free MVNs connectivity prediction platform. Firstly, it is possible to independently cultivate mature, perfusable MVNs in a microfluidic device, as well as to simulate the growth of MVNs within the tumor microenvironment. To study the connectivity of these MVNs, we used MVNs bright field images as the raw images and the post-perfusion fluorescent images as the labels to form an MVNs connectivity prediction dataset. Subsequently, we proposed a deep learning method based on the contrastive learning and a class imbalanced algorithm, designated as the Vessel Connectivity Network (VC-Net), for the segmentation of the morphology of in vitro vascular models with perfusion functionality and the achievement of label-free connectivity prediction. This network is able to train with high accuracy using only a small number of data samples, thus reducing the high costs associated with conducting excessive repeated experiments to meet training sample size requirements. Moreover, we propose a class imbalance algorithm to enhance the overall performance and robustness of the VC-Net, addressing the problem of severe class imbalance in the dataset. To our knowledge, our team is the first to develop a deep learning-based, label-free MVNs connectivity prediction platform. The subsequent culture process of vascularized organoids or tumors on our platform obviates the need to perfuse fluorescent dyes or microspheres for the purpose of MVNs connectivity. The trained VC-Net enables direct prediction, thereby reducing the cost and time associated with the culture process. This platform could provide insights into whether MVNs provide nutrients and remove metabolic waste during organoid development or tumor progression under various physiological and pathological conditions.

## Methodology

### Establishment of a Label-free MVNs Connectivity Prediction Platform

To culture mature, perfusable MVNs, inspired by previous studies,[8, 10, 39, 40] the microfluidic device was designed with alternating gel and medium channels, separated by parallel micro-pillars spaced 100 μm apart to prevent the hydrogel from rupturing into the medium channels (illustrated in **Fig. 1**a). Human umbilical vein endothelial cells (HUVECs) were encapsulated within the central gel region (HUVECs-gel channel D), while Human lung fibroblasts (FBs) were encapsulated within the adjacent gel regions on either side (FBs-gel channels B and F). The medium channels (medium channels A, C, E, and G) were filled with cell culture medium, resulting in perfusable MVNs after 6 days (**Fig. 1**b).

In order to establish a label-free MVNs connectivity prediction platform, it was first necessary to create a perfusion connectivity dataset for MVNs. The perfusion experiment entailed the injection of a fluorescent dye into medium channel C, allowing it to flow through the MVNs. The dye then flowed out of the culture medium channel E (as depicted in **Fig. 1**c-i). Paired bright-field and fluorescent perfusion images of the same field of view were collected using confocal microscopy.

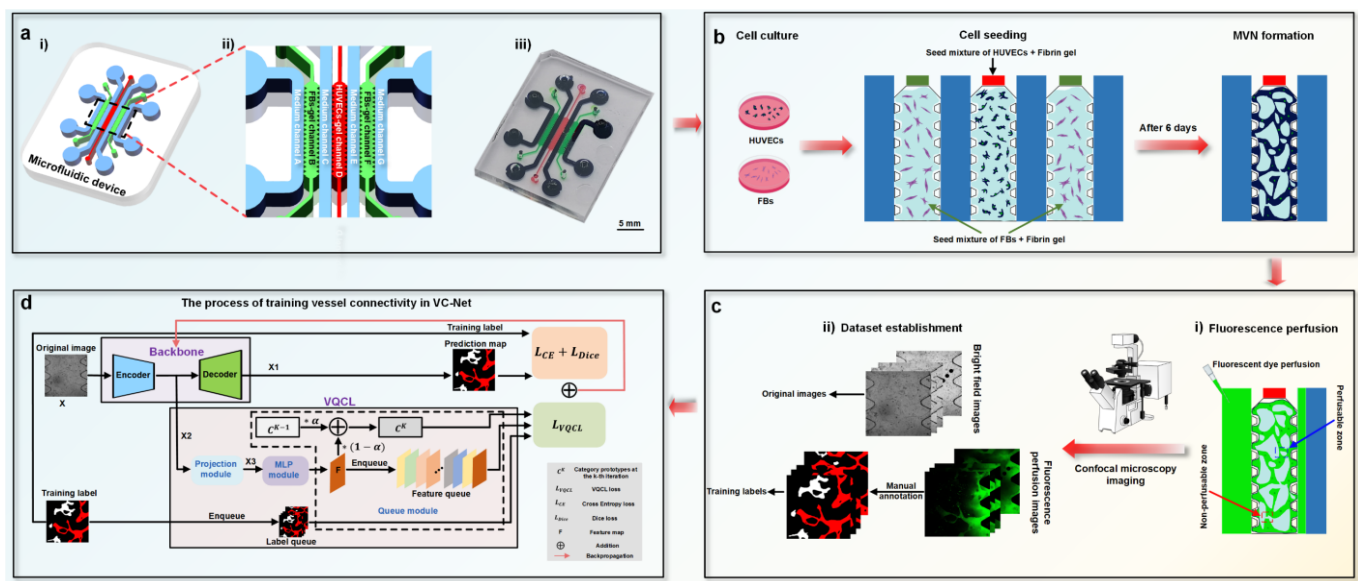

**Fig. 1** The overall process of establishing the label-free MVNs connectivity prediction platform based on VC-Net. **a)** Schematic diagram of the microfluidic device, an enlarged view, and a photograph of the device filled with colored hydrogel. **b)** Experimental setup and timeline for MVNs formation. **c)** i) Fluorescence perfusion of MVNs; ii) Generation of a small-scale image dataset (bright-field, fluorescence images, and labeled images). **d)** The overall workflow of training VC-Net for predicting MVNs connectivity, including the network's backbone (encoder and decoder), VQCL method, and loss function backpropagation.

Furthermore, to facilitate model training, the fluorescent perfusion images were binarized to generate dataset labels (**Fig. 1**c-ii). In these labels, red indicates perfused, connected regions, white denotes non-connected regions, and black represents the background. Similarly, experiments were conducted on the perfusion of MVNs within the tumor microenvironment, resulting in datasets on the connectivity of MVNs under these conditions. This approach facilitated the development of a platform for the prediction of label-free MVNs connectivity in tumor cultures.

Finally, we propose a deep learning network, the VC-Net, which enables label-free prediction of MVNs connectivity. The bright-field images of the MVNs are utilized as inputs to the network model, which are processed by the encoder to extract semantic features and by the decoder to generate the prediction map, thereby extracting MVNs structural features (**Fig. 1**d). The prediction map is then compared with the training label (ground truth) to calculate the combined loss of Cross Entropy (CE) loss [41] and Dice loss.[42] This is done in order to achieve better performance and robustness. A novel VQCL method is introduced in VC-Net to effectively determine the connectivity of MVNs by increasing the difference between categories. In addition, to address the issue of class imbalanced in the connectivity dataset, we introduce a class imbalanced algorithm in CE loss. Subsequently, the overall loss value is calculated and propagated backwards through the network to optimize the parameters of the model. The training process is repeated multiple times until the predefined number of training epochs has been reached. The trained VC-Net obviates the necessity for fluorescent perfusion, as it is capable of generating prediction maps that are highly analogous to the ground truth by inputting bright-field images of MVNs from the test set. This enables the assessment of whether MVNs contribute to nutrient supply and waste metabolism during organoid development or tumor progression.

**VC-Net Architecture**

We propose a network, the VC-Net, for the objective assessment of the connectivity within in vitro MVNs models with perfusion functionality. The overall framework of VC-Net comprises three principal components: an encoder-decoder-based backbone, the VQCL method, and a class imbalanced algorithm, as illustrated in **Fig. 1**d and **2**a. For each image $x \in \mathbb{R}^{H \times W \times 3}$ in the MVNs connectivity dataset, the training process involves the extraction of semantic features from the input image using the encoder in the backbone. The decoder then restores these features to the original image resolution to generate a predicted map $x_1 \in \mathbb{R}^{H \times W \times L}$ (with a total of $L = 3$ classes: connected, non-connected, and background classes). In order to obtain a more precise MVNs skeleton structure, we selected and compared various networks based on U-Net and DeepLab as the backbone for VC-Net. The predicted map $x_1$ is compared with the ground truth, and the loss value is calculated using a combined loss function that combines the commonly used CE loss and Dice loss in segmentation tasks. The model parameters are updated continuously by computing the gradient of the loss function and performing backpropagation until the training is complete, resulting in the optimal model parameters.

Furthermore, we propose a VQCL method, which comprises three independent modules: the Projection module, the MLP module, and the Queue module, as illustrated in **Fig. 2**. This method employs the use of class center features and features of images from the same class as positive sample pairs, along with features of images from different classes as negative sample pairs for contrastive learning. This approach facilitates more effective differentiation of the various classes within the MVNs connectivity dataset. Furthermore, to meet the sample size requirements for training while reducing the cost of excessive repetitive experiments, VQCL employs a Queue module to store the features and corresponding labels from previous iterations. These stored features and labels are also employed in contrastive learning during training, thereby augmenting the sample data and enhancing the model's robustness.

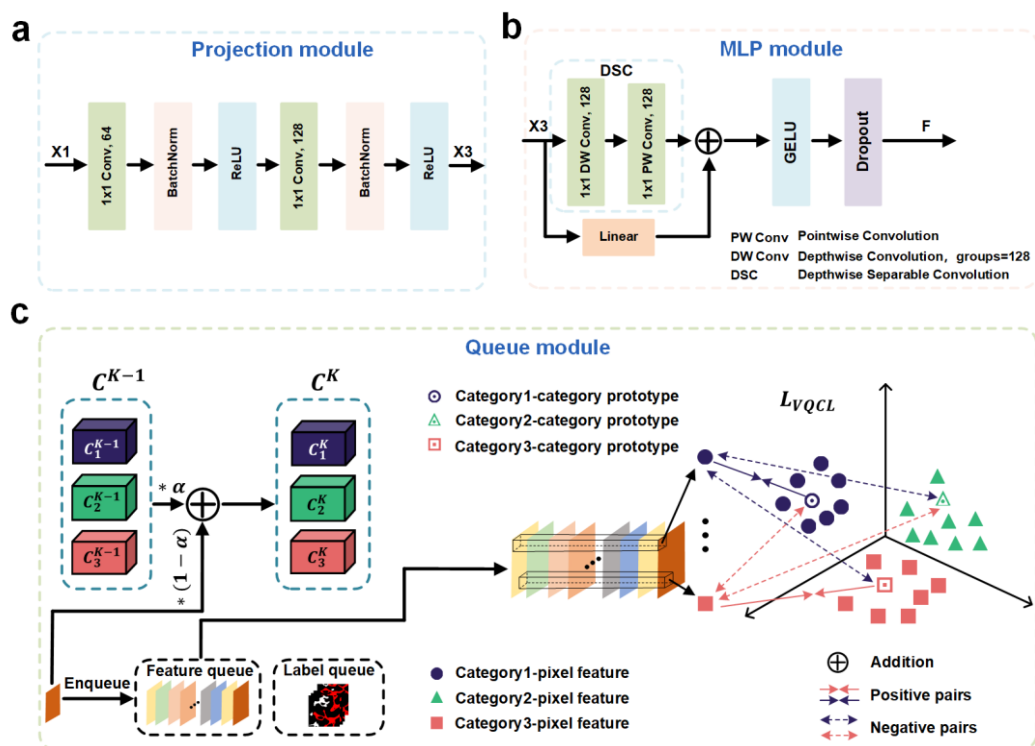

**Fig. 2 The overall structure of the proposed VQCL in VC-Net. a)** The architecture of the Projection module in VQCL. **b)** The architecture of the MLP module in VQCL. **c)** The architecture of the Queue module in VQCL.

Finally, to address the issue of class imbalanced in MVNs connectivity analysis, which leads to the non-connected class being incorrectly classified as connected or background classes, we introduced and compared several methods for handling class imbalanced. These methods include re-weighting (RW),[43, 44] deferred re-weighting (DRW),[43] and our previous work on phased progressive weighting (PPW).[45] The aforementioned methods enhance the weight of minority samples (such as non-connected classes) by incorporating a loss weighting factor that inversely alters the number of classes in the loss function. This results in the network directing greater attention to these minority classes.

Vessel Queue Contrastive Learning

During network training, the high-level semantic features extracted by the encoder designed for image segmentation tasks (**Fig. 1**d) frequently contain a substantial number of redundant features. This redundancy can constrain the representational capacity of the features and diminish the distinction between classes in feature distribution.[23, 46] In recent years, contrastive learning has been extensively incorporated into the training of deep learning models with the objective of enhancing their representation learning capabilities.[47-53] However, when the number of samples available for training is insufficient, the ability of contrastive learning to distinguish between different classes is limited.

To address these issues, we propose a VQCL method, which comprises three independent modules: the Projection module, the MLP module, and the Queue module, as illustrated in **Fig. 2**. The VQCL sequentially maps features to a lower-dimensional embedding space through the Projection module (**Fig. 2**a) and the MLP module (**Fig. 2**b), resulting in the feature F. The Projection module reduces the number of channels in the original feature F from 2048 to 128 by introducing two 1x1 convolutions, denoted as $x_3 \in \mathbb{R}^{\frac{H}{16} \times \frac{W}{16} \times 128}$, while maintaining the feature size unchanged. The formula for the Projection module is as follows:

$$x_2' = ReLU(BN(Conv_{1\times1}(x_2))) \quad (1)$$

$$x_3 = ReLU(BN(Conv_{1\times1}(x_2'))) \quad (2)$$

where $BN(.)$ denotes the BatchNorm technique,[54] while $ReLU(.)$ denotes the ReLU activation function.[55]

Subsequently, the MLP module employs a 1x1 Depthwise Convolution (DW Conv)[56] to reduce the number of parameters in the convolutional layer, thereby decreasing computational complexity. This is followed by a 1x1 Pointwise Convolution (PW

Conv),[56] which enhances the feature representation capability. Concurrently, a skip connection *Linear(.)*[57] is employed to enhance the network's capacity for generalization. The DW Conv and PW Conv are collectively designated as Depthwise Separable Convolution (DSC).[58] The formula for the MLP module is as follows:

$$x_3' = PWConv_{1\times1}(DWConv_{1\times1}(x_3)) + Linear(x_3) \quad (3)$$

$$F = GELU(Dropout(x_3')) \quad (4)$$

where *Dropout(.)* is a regularization technique used to prevent overfitting,[59] while *GELU(.)* denotes the GELU activation function.[60]

The features extracted by the Projection and MLP modules will have higher quality. To further optimize the efficacy of contrastive learning, a class centers-based contrastive learning method is employed. In particular, for each class within the dataset, the class center features $C^{(0)} = \{C_0^{(0)}\ C_1^{(0)}\ C_2^{(0)}\}$ are randomly initialized. Subsequently, to ensure that the class centers increasingly accurately reflect the feature information of each class during model training, we progressively update the class centers in each iteration, as illustrated in **Fig. 2**c. Specifically, for the *K*-th iteration, the class center $C_i^{(K)}$ for class *i* can be updated based on the pixel features of that class in each batch as follows:

$$C_i^{(K)} \leftarrow \alpha C_i^{(K-1)} + (1-\alpha)\frac{1}{|B_i|}\sum_{b=1}^{B}\sum_{m}^{H\times W} y_{i,b}[m] f_b^{(K)}[m] \quad (5)$$

where *B* represents the batch size, $|B_i|$ denotes the number of pixels belonging to class i in the batch size, and m represents any pixel in images. $y_{i,b}[m]$ is the label of pixel m for class i in the *b*-th sample ( $y \in [0,1,2]$ ), and $f_b^{(K)}[m]$ is the feature of pixel *m* at the *K*-th iteration, corresponding to the feature F in Equation 4. *α* is the momentum hyperparameter for updating the class centers, empirically set to 0.4. With continuous training, the accuracy of the class center for each class is progressively optimized.

In order to address the issue of a severe shortage of training samples, a Queue module has been incorporated into VQCL. This module stores the features obtained from each computation in the Queue. VQCL considers the class center of each pixel belonging to a specific class and its corresponding feature in the Queue (e.g., the feature of class i in the *K*-th iteration) as positive sample pairs $(C_i^{(K)}, f_i^+)$. Concurrently, the class center is paired with features of other classes in the Queue as negative sample pairs $(C_i^{(K)}, f_l^-)(l \neq i)$. This method markedly amplifies the number of positive and negative sample pairs, thereby facilitating more accurate representation learning. Accordingly, the VQCL loss for class *i* in the *K*-th iteration ($\mathcal{L}_{VQCL}^i$) and the total VQCL loss ($\mathcal{L}_{VQCL}$) are formulated based on the popular contrastive loss InfoNCE,[61, 62] as follows:

$$\mathcal{L}_{VQCL}^i = -\mathbb{E}_{f_i^+ \in Q} \log \frac{e^{[C_i^{(K)}]^T f_i^+/\tau}}{e^{[C_i^{(K)}]^T f_i^+/\tau} + \sum_{l \neq i}^{L-1} \mathbb{E}_{f_l^- \in Q} e^{[C_i^{(K)}]^T f_l^-/\tau}} \quad (6)$$

$$\mathcal{L}_{VQCL} = \sum_{i=0}^{L-1} \mathcal{L}_{VQCL}^i \quad (7)$$

where $f_i^+$ represents the feature of the positive sample of class i extracted from the positive sample of the Queue, and $f_l^-$ represents the feature of other classes l extracted from the negative sample of the Queue. $\mathbb{E}[.]$ denotes the averaging operation, aiming to normalize the samples onto the unit sphere. $\tau$ is the temperature hyperparameter controlling the gradient penalty for positive and negative samples, empirically set to 0.4. During the iteration process, the Queue adheres to the principle of first-in, first-out, whereby the most recent batch is added to the Queue while the oldest batch is removed. The size of the queue is set to 128.

Loss Function

To further improve the accuracy of connectivity prediction, in addition to using the contrastive learning loss function, we have integrated a combined loss function commonly employed in segmentation: CE loss[41] and Dice loss.[42]

The CE loss is employed to direct the model to accurately classify each pixel by measuring the discrepancy between the predicted probability distribution and the true label distribution. The formula is as follows:

$$\mathcal{L}_{CE} = -\log(p_y) \quad (8)$$

where $p_y = e^{z_y} / \left(\sum_{j=1}^{L} e^{z_j}\right)$, L represents the total number of classes. $z_y$ denotes the predicted output for the true label y of the pixel, and $z_j$ denotes the predicted output for the pixel's labels of other classes.

Dice loss is one of the most prevalent loss functions in the field of image segmentation. The measure of overlap between the predicted region and the ground truth region is employed. The formula is as follows:

$$\mathcal{L}_{Dice} = 1 - Dice = 1 - \frac{2*|X \cap Y|}{|X| + |Y|} \quad (9)$$

where X and Y represent the binary masks of the truth labels and the predicted results, respectively.

Finally, the CE loss, Dice loss, and VQCL loss are combined to form the final segmentation loss, which quantifies the differences between the ground truth and the model predictions. The combined loss function is formulated as follows:

$$\mathcal{L}_{total} = \mathcal{L}_{CE} + \mathcal{L}_{Dice} + \mathcal{L}_{VQCL} \quad (10)$$

Class Imbalanced Algorithm

To address the class imbalanced issue, we can apply reweighting to the CE loss, which is a widely used method. The reweighting (RW) method is a common method employed in the field of imbalanced visual recognition,[43, 44] typically introducing a loss weight factor inversely proportional to the number of samples in the loss function:

$$\mathcal{L}_{CE} = -\left(\frac{1}{n_y}\right)\log(p_y) \quad (11)$$

where n is the number of pixels in the class labeled y.

Cao et al.[43] initially proposed the DRW method, which involves first performing regular training and then training with the RW method in a second stage. The training in stage 1 serves as an effective foundation for the training in stage 2. Our previous work introduced a highly effective method for addressing class imbalanced, called PPW.[45] This method introduces a gradual transition phase to the DRW method, facilitating a smoother progression from general representation learning to classifier training. Consequently, PPW modifies the weights of the loss function in **Equation 11** as follows:

$$\omega_i = \left(\frac{1}{n_i}\right)^\beta \quad (12)$$

where $n_i$ is the number of pixels in class i, $\omega_i$ is the loss weight for class i, and $\beta$ is formulated as follows:

$$\beta = \begin{cases} 0, & E < E_{min} \\ \left(\frac{E - E_{min}}{E_{max} - E_{min}}\right)^2, & E_{min} \leq E \leq E_{max} \\ 1, & E > E_{max} \end{cases} \quad (13)$$

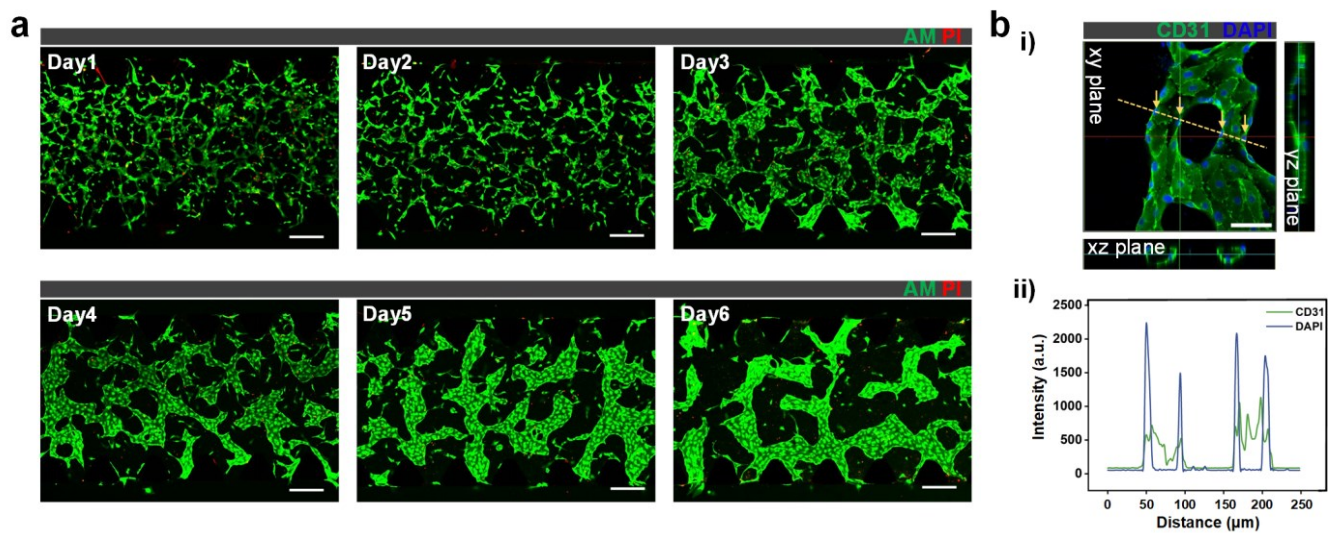

**Fig. 3 The formation and parameter measurement of MVNs in microchannels. a)** Confocal images of HUVECs stained with AM/PI from Day 1 to Day 6 post-seeding, illustrating the development of MVNs at different time points, with live cells (green) and dead cells (red). Scale bar: 200 μm. **b)** i) Confocal images of MVNs in frontal view and cross-sections, showing MVNs lumens, the images display a line scan measurement (yellow line); ii) fluorescence intensity histograms of MVNs along the scanned region. CD31 (green) and DAPI (blue). Scale bar: 50 μm.

In the initial stage of training ($E < E_{min}$), each class is assigned the same loss weighting factor ($\beta = 0$, $\omega_i = 1$). During the progressive transition phase ($E_{min} \leq E \leq E_{max}$), the value of $\beta$ smoothly varies from 0 to 1 over the epochs. Similarly, in the final stage ($E > E_{max}$), the weights are set to values inversely proportional to the number of samples in each class ($\beta = 1$, $\omega_i = (1/n_i)$), thereby increasing the training weight for minority samples classes. We empirically choose the PPW parameters $E_{min} = 100$ and $E_{max} = 200$, allowing the model to quickly adjust weights according to the differences in class quantities to minimize loss.

## Results

### The Formation of Perfusable MVNs
During the cultivation process, HUVECs undergo a simulated vascular self-assembly process. After one day, they form vascular fragments, and gradually develop branches. After six days, they grow into a mammalian-like interconnected MVNs, as shown in **Fig. 3**a, **Fig. S1**. For the mature MVNs, CD31 and DAPI are employed to specifically label HUVEC morphology and cell nuclei. The measurements and statistics of MVNs parameters during the growth process (Day 2 to Day 6). The average diameter of the MVNs increased from 42 μm to 70 μm, the total area of the MVNs grew from 2.3 mm² to 3.1 mm², and the total length of the MVNs extended from 42.5 mm to 48 mm, which is consistent with related literature reports.[18, 24, 63] Simultaneously, Cell Counting Kit-8 (CCK-8) assays were conducted from Day 1 to Day 6 (**Fig. S2**a-b). The absorbance (OD) of the culture supernatant was quantified using a microplate reader to assess the proliferation and viability of cells in the MVNs. As illustrated in **Fig. S2**c, the proliferation rate of cells significantly increased from Day 1 to Day 4 and continued to proliferate steadily after Day 5. Additionally, the utilisation of confocal microscopy with 3D scanning functionality enables the acquisition of both frontal and cross-sectional views of the MVN (**Fig. 3**b-i), as well as the tracking of the fluorescence intensity distribution of CD31 and DAPI in HUVECs (**Fig. 3**b-ii). The observation results demonstrated that the luminal structure was transparent in the xz and yz planes, indicating that the cultured MVN was perfusable.

### Comparison Results of Backbone

In recent years, networks such as U-Net and DeepLab have been widely used for image segmentation in biomedical and other fields.[36, 42, 64, 65] The principal advantage of these networks lies in their use of an encoder-decoder architecture, which enables the effective extraction of features and the recovery of structural information in images. The encoder employs a series of convolutional and pooling operations to gradually extract the image's deeper features. Subsequently, the decoder progressively restores the image's detail information through upsampling and convolution operations, thereby achieving pixel-level precise segmentation. Consequently, the utilization of architectures such as U-Net or DeepLab as the encoder and decoder of VC-Net (**Fig. 1**d), that is, the backbone of VC-Net, can effectively extract information about the overall structure and edges of the MVNs.

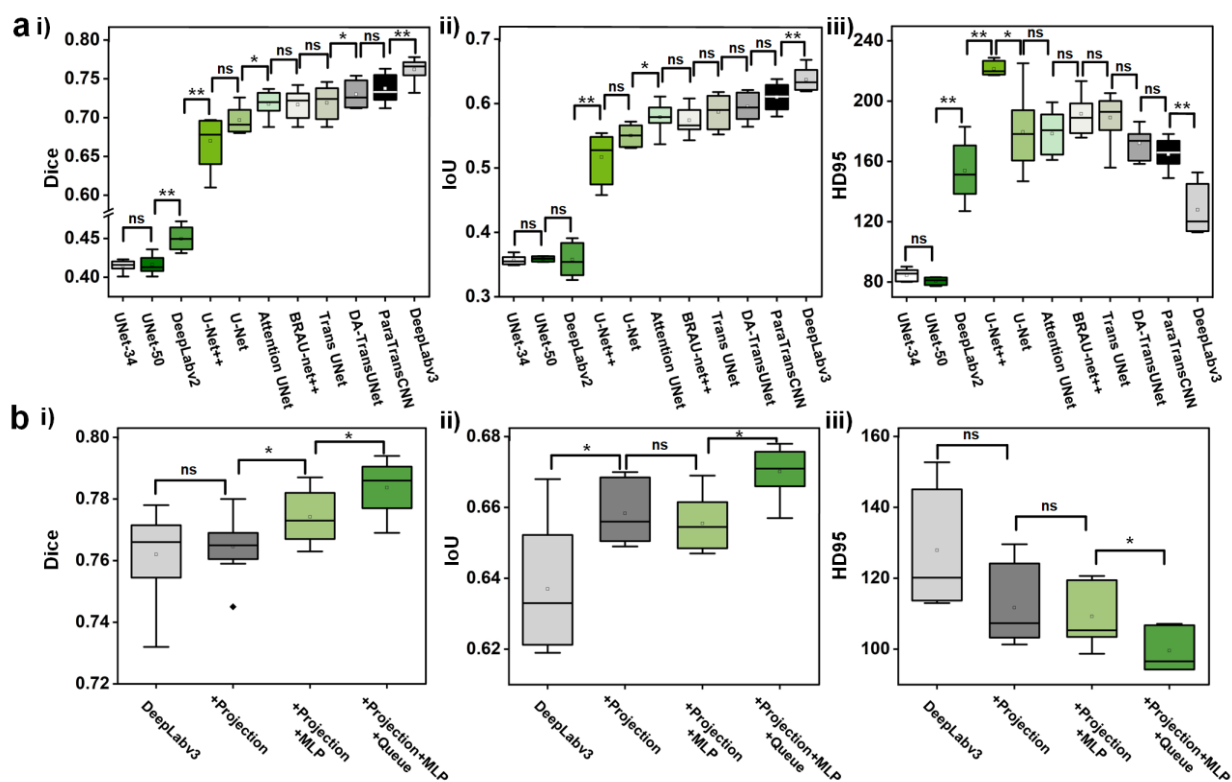

**Fig. 4 Performance comparison of backbones and contrastive learning modules in VC-Net. a)** Comparison of different backbones on the connectivity dataset in terms of i) Dice, ii) IoU, and iii) HD95 metrics (n=10). **b)** Performance comparison of introducing various contrastive learning modules on the connectivity dataset in terms of i) Dice, ii) IoU, and iii) HD95 metrics (n=10). The error bars represent the standard deviation (SD), and the small squares indicate the mean, with the box lines indicating the 25%, 50%, and 75% percentiles from the bottom to the top.

In order to identify the optimal backbone, a comparative analysis was conducted of the performance of various backbones on the MVNs connectivity dataset. The following backbones were considered: U-Net,[42] U-Net++,[66] U-Net based on ResNet-34 (UNet-34) and ResNet-50 (UNet-50).[67] In addition, advanced high-precision Transformer methods have been implemented in recent years, such as Attention Unet[68], BRAU-net++[69], TransUNet,[70] DA-TransUNet[71] and ParaTransCNN[72] as well as DeepLab architectures, including DeepLabv2[64] and DeepLabv3.[65] **Fig. 4**a illustrates the outcomes of diverse backbones in establishing connectivity on the test set, evaluated through the Dice, intersection over union (IoU), and HD95 metrics.

Although UNet-34 and UNet-50 demonstrate superior performance relative to other methods in the HD95 metric, they exhibit a notable discrepancy in Dice and IoU values when compared to the other methods. Additionally, the visualization of the prediction outcomes for all backbones is presented in **Fig. 5**, UNet-34 is unable to identify non-connected class. Consequently, in order to select the most appropriate backbone, UNet-34 and UNet-50 are initially eliminated from consideration. Among the remaining networks, it was found that DeepLabv3 exhibited the most optimal performance. Not only does it exhibit the highest average Dice and IoU values (exceeding the second-best ParaTransCNN by 0.025 and 0.027, respectively), but it also has the lowest average HD95 value (25.98 lower than the second-lowest DeepLabv2). Therefore, we selected DeepLabv3 as our backbone. In **Fig. S3**a, we also compared the performance of different backbones in terms of Average Surface Distance (ASD) and Accuracy (Acc), with the results indicating that DeepLabv3 exhibited the greatest efficacy.

Results of VQCL

Although introducing a backbone with excellent performance is an effective method for identifying the overall structure of MVNs, the differentiation between connected and non-connected class remains a challenge. A reliance on the backbone alone is inadequate for the analysis of MVNs connectivity. The VQCL method is integrated into VC-Net with the objective of enhancing the precision of MVNs connectivity prediction through the introduction of contrastive learning. To assess the efficacy of VQCL on the MVNs connectivity dataset, a comparison was conducted with the original DeepLabv3. Following the incorporation of the Projection module, VQCL demonstrated a 0.02 improvement in IoU, with no discernible alterations in Dice and HD95, as illustrated in **Fig. 4**b. Following the introduction of both the Projection module and the MLP module, there was a 0.01 improvement in Dice, a 0.02 improvement in IoU, and a 18.63 reduction in HD95. This indicates that finer feature representations enhance the effectiveness of contrastive learning. Finally, the introduction of the Queue module on this basis resulted in a 0.02 improvement in Dice, a 0.03 improvement in IoU, and a 28.31 reduction in HD95 compared to DeepLabv3. This demonstrates that the proposed VQCL's Projection module, MLP module, and Queue module can all effectively enhance the discriminative ability for MVNs connectivity. As illustrated in **Fig. 5**, the introduction of VQCL greatly improves the ability to distinguish classes information. Similarly, in **Fig. S3**b, we demonstrate the efficacy of VQCL in ASD and Acc, with the results indicating its superiority.

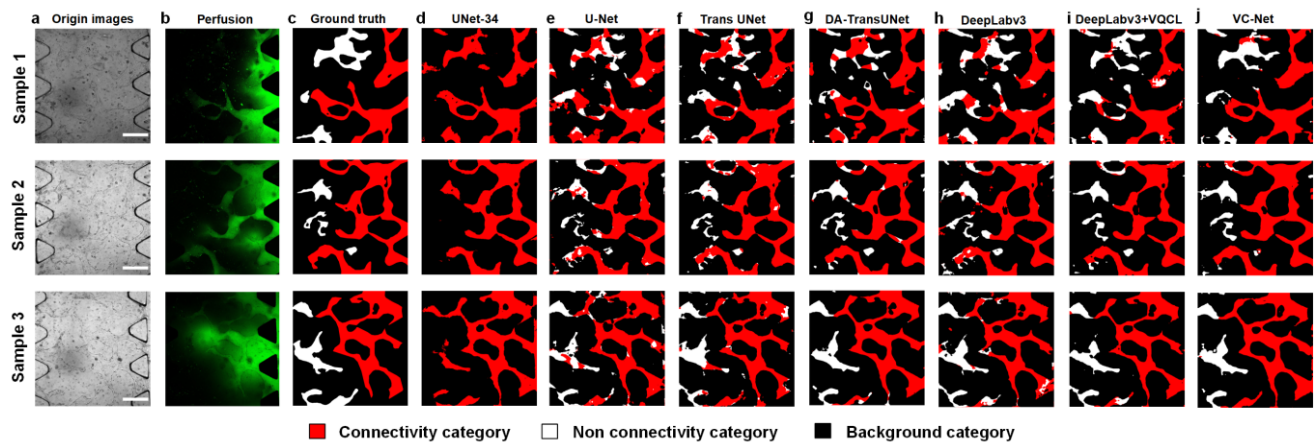

**Fig. 5 A qualitative comparison of different backbones and VC-Net is performed through visualization on the MVNs connectivity dataset.** The images are presented from left to right, as follows: **a)** origin images, **b)** fluorescent perfusion images, **c)** ground truth, **d)** UNet-34, **e)** U-Net, **f)** Trans UNet, **g)** DA-T Trans UNet, **h)** DeepLabv3, **i)** DeepLabv3+VQCL, and **j)** VC-Net. The color red indicates the connected class, the color white indicates the non-connected class, and the color black indicates the background class. Scale bar: 200 µm.

Results of Class Imbalanced Algorithm

Despite the efficacy of the Deeplabv3+VQCL method in segmenting and discriminating between the morphological and connectivity classes of MVNs, instances of misclassification persist, with non-connected class being erroneously identified as connected or background classes. This may be attributed to the imbalanced distribution of samples across the different classes in the dataset. To validate our hypothesis, we calculated the area proportions of the three classes in the MVNs connectivity dataset, as illustrated in **Fig. 6**a. In the test set, the average area proportion for the background class is 62.7%, for the connected class is 29.9%, and for the non-connected class is 7.4%. This indicates that the dataset exhibits a severe class imbalanced issue. Due to the smaller number of pixels for the non-connected class, the model may be biased towards the majority classes during training, preventing the network from adequately learning the features of these minority classes. Furthermore, the loss function values of VQCL were recorded during the network training process and the total loss function values were recorded after the introduction of PPW. The changes in these values across epochs are shown in **Fig**. **6**b. The two loss curves demonstrate a gradual decrease in loss values as the number of epochs increases. This indicates that the model is progressively learning to better predict the target variables, thereby reducing prediction error. In the early stages of training, the model's randomly initialized weights are typically far from the optimal solution. The interval where the loss value of both methods decreases fastest is between 0 and 200 epochs, so the interval of the progressive transition phase are selected as $E_{min}=100$ and $E_{max}=200$.

In order to more accurately distinguish the minority non-connected class, we conducted a comparative analysis of the performance of class imbalanced algorithms on the MVNs connectivity dataset. This analysis was based on the encoder-decoder and VQCL (DeepLabv3+VQCL) foundations. As illustrated in **Fig. 6**c, the RW method demonstrated no significant differences in Dice, IoU, and HD95 compared to the DeepLabv3+VQCL method. Nevertheless, the DRW method exhibited a 0.02 increase in the average Dice score, a 0.01 rise in the average IoU, and a 5.45 reduction in the average HD95. This suggests that the incorporation of the DRW method enhances the network's capacity to predict classes and boundaries. In comparison to the DRW method, the PPW method demonstrated no significant difference in average HD95 (86.33), yet exhibited a notable increase in average Dice (0.81) and IoU (0.70). This evidence suggests that PPW exhibits the most effective performance in addressing the class imbalanced problem in MVNs connectivity. Consequently, the PPW method was integrated into VC-Net. As illustrated in **Fig. 4**i, the incorporation of class imbalanced methods into VC-Net enhances the accuracy of distinguishing the non-connected class without compromising the prediction performance for the connected and background classes. Similarly, in **Fig. S3**c illustrates the performance of the PPW in ASD and Acc. The results indicated the superiority of PPW.

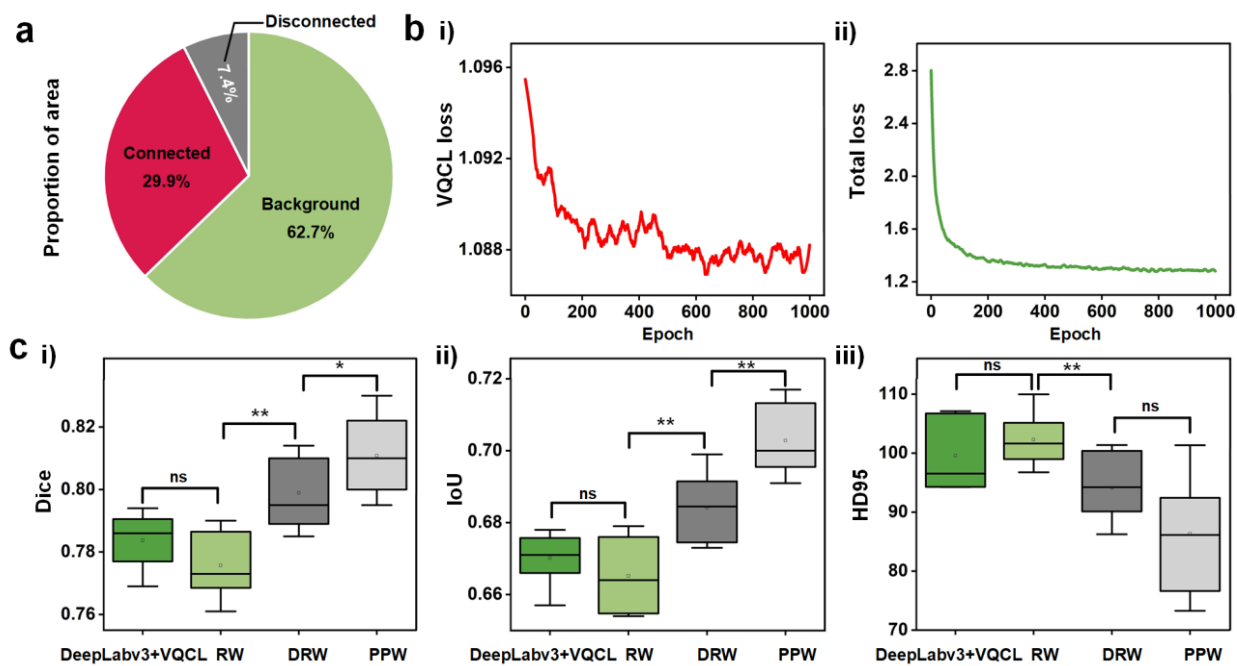

**Fig. 6** Performance comparison of class imbalanced algorithms in VC-Net. **a)** Area proportions of the three classes in the test set of the MVNs connectivity dataset (n=3). **b)** Variation curves of i) VQCL loss and ii) total loss values with epochs in VC-Net. **c)** Performance comparison of class imbalanced methods RW, DRW, and PPW, introduced on top of DeepLabv3+VQCL, in the MVNs connectivity dataset for i) Dice, ii) IoU, and iii) HD95 metrics (n=10). The error bars represent the standard deviation (SD), and the small squares indicate the mean, with the box lines indicating the 25%, 50%, and 75% percentiles from the bottom to the top.

**Analysis of MVNs Connectivity in Tumor Microenvironment**

The experimental results demonstrate that our VC-Net exhibits excellent performance in conventional MVNs culture detection. To further validate the generalizability of our label-free MVNs connectivity prediction platform and the VC-Net in tumor cultures, we also analyzed the impact of MVNs connectivity within the tumor microenvironment as an example.[73, 74] Hu et al.[15] found that in a chip-based co-culture of blood vessels and tumors, the blood vessels in the tumor-surrounding areas had smaller diameters but a denser vascular network compared to vessels in other regions. The co-culturing of MVNs with the medium derived from tumor cell cultures has been demonstrated to induce the characteristics of tumor-associated vasculature without the need for direct incorporation of tumor cells. This process results in the formation of a consistently disordered and twisted vascular network.[75] To comprehensively examine the connectivity of MVNs in a tumor microenvironment, we utilized a 1:1 mixture of tumor cell medium (TCM) derived from human liver cancer cells (HepG2s) culture and EGM-2 to simulate the tumor microenvironment (**Fig. S**4). **Fig. 7**a presents confocal images and detailed images of MVNs cultured under EGM-2 and TCM ([1:1] EGM-2: TCM) on Days 1, 3, and 6. It can be observed that MVNs cultured with TCM exhibit decreased diameter and increased density and complexity of the network. **Fig. 7**b presents a comparison of the MVNs parameters under the two conditions. There was no significant difference in the average area of MVNs in EGM-2 and TCM under the same cell concentration. This may be attributed to the fact that the initial cell concentration and cell proliferation were consistent under the two conditions (**Fig. S**5). Furthermore, the findings indicated that the average diameter of MVNs in EGM-2 and TCM is 70 μm and 54 μm, respectively, while the average total length is 48 mm and 56 mm, respectively. In comparison to conventional culture, the diameter of MVNs in the tumor microenvironment is significantly reduced, and a greater number of disordered vascular branches are produced, resulting in a significant increase in total vessel length, which is similar to the results of previous studies.[15, 73, 74, 76]

Subsequently, the performance of VC-Net and various deep learning networks in predicting the connectivity of MVNs in the tumor microenvironment was evaluated. The results are presented in **Fig. 8**a-b. The UNet-34 model is still unable to accurately identify the morphology of MVNs, and it is unable to determine the connectivity of MVNs. In comparison to the original U-Net (average Dice of 0.61 and average HD95 of 294.77), ParaTransCNN demonstrated no significant difference in Dice score, yet exhibited a reduction in HD95 by 135.08. In contrast, the VC-Net model demonstrated an improvement in the Dice score by 0.06, reaching a value of 0.71. Moreover, the HD95 metric was reduced by 63.88, resulting in a value of 95.81. **Fig. 8**a illustrates that VC-Net outperforms other methods in both the overall shape segmentation of MVNs and the prediction of connectivity. Similarly, in **Fig. S**6, we conducted performance comparisons based on IoU, ASD, and Acc, with the results indicating the superiority of VC-Net.

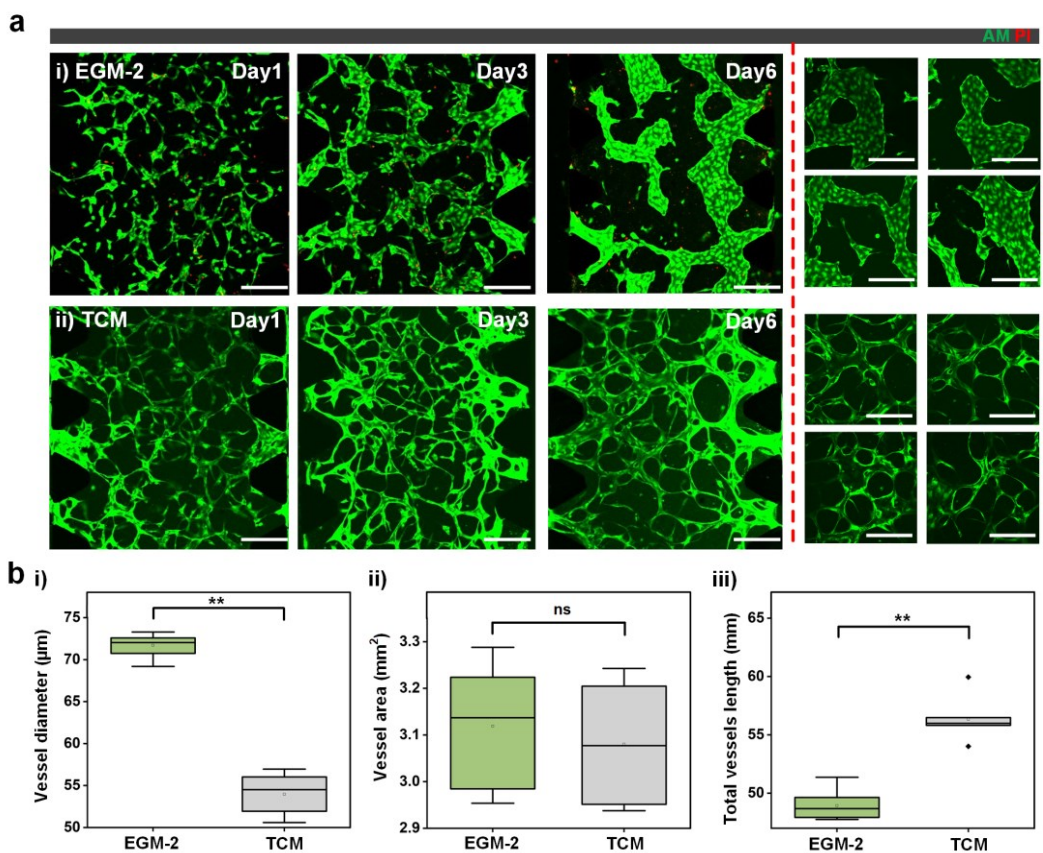

**Fig. 7 MVNs cultured with EGM-2 and TCM. a)** Fluorescence images of self-assembled, fully perfusable vascular networks obtained on Days 1, 3, and 6 post-seeding (left, scale bar: 200 μm) and detail images (right, scale bar: 200 μm). i) MVNs cultured with EGM-2 medium; ii) MVNs cultured with [1:1] EGM-2: TCM. **b)** A comparison was conducted between MVNs cultured with EGM-2 and TCM on Day 6 for i) vessel diameter (μm), ii) Vessel area (mm2), and iii) total vessel length (mm) (n=8). The error bars represent the standard deviation (SD), and the small squares indicate the mean, with the box lines indicating the 25%, 50%, and 75% percentiles from the bottom to the top.

Finally, we compared the proportions of each class of MVNs under EGM-2 and TCM culture conditions in the test set with the predictions made by VC-Net, as shown in **Fig. 8**c. Under TCM culture conditions, the average proportion of the connected class is 20.7%, while the average proportion of the non-connected class is 11.8%. Compared to EGM-2 culture, the average number of the connected class under TCM culture conditions decreases by 30.8%, while the average number of the non-connected class increases by 37.3%. This indicates that the connectivity of MVNs changes under the influence of the tumor microenvironment, with some vascular channels becoming narrower and impeding fluid flow, leading to a reduction in connected areas and an increase in non-connected areas. This finding is consistent with previous research,[76-80] which indicates that tumor vasculature exhibits abnormal blood flow, which can lead to occlusion in surrounding vessels and subsequent impairment of MVNs connectivity. A comparison of the VC-Net results with the ground truth data obtained under EGM-2 and TCM culture conditions reveals no significant differences across the three classes. This indicates that the predictions made by VC-Net are highly consistent with the observed connectivity in the ground truth data obtained through fluorescence perfusion, irrespective of whether the conditions are EGM-2 or TCM.

## Conclusions and Discussion

In the field of tissue engineering, it has been demonstrated that the connectivity of MVNs cultured in vitro affects the development of co-cultured organoids or tumors. However, this connectivity is often heavily dependent on fluorescent labeling methods, which may hinder the further maturation of organoids and tumor spheres. This study presents a deep learning-based label-free MVNs connectivity prediction platform. Initially, MVNs were constructed under conventional culture and tumor microenvironment conditions in a microfluidic device. The original images of MVNs were utilized as training samples, while perfused fluorescent images served

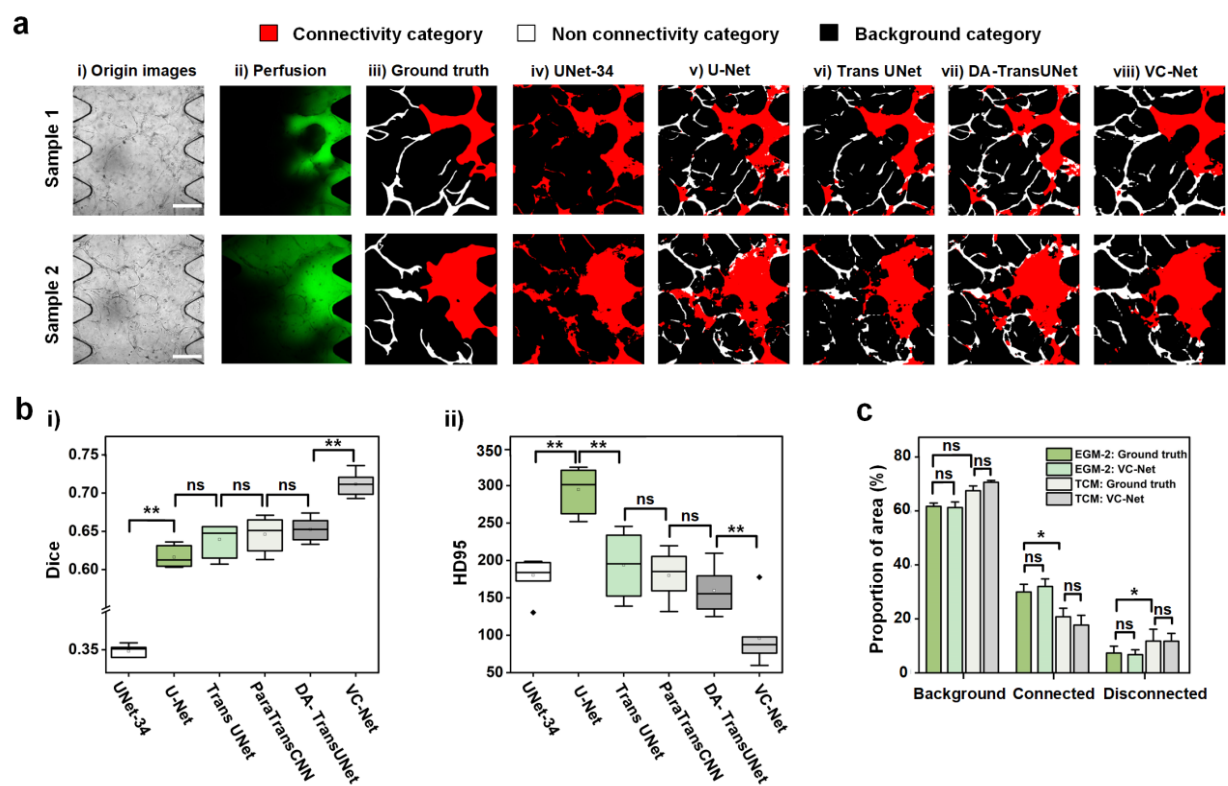

**Fig. 8** VC-Net connectivity prediction analysis for MVNs cultured in TCM. **a)** A qualitative comparison of different backbones and VC-Net is performed through visualization on the TCM-cultured MVNs connectivity dataset. The images are presented from left to right, as follows: i) origin images, ii) fluorescent perfusion images, iii) ground truth, iv) UNet-34, v) U-Net, vi) Trans UNet, vii) DA- Trans UNet, and viiI) VC-Net. The color red indicates the connected class, the color white indicates the non-connected class, and the color black indicates the background class. Scale bar: 200 μm. **b)** Comparison of different backbones and VC-Net on the TCM-cultured MVNs connectivity dataset in terms of i) Dice, ii) HD95 metrics (n=10). **c)** A comparison of the area proportions of three classes in MVNs test sets under EGM-2 and TCM culture conditions (n=7). The error bars represent the standard deviation (SD), and the small squares indicate the mean, with the box lines indicating the 25%, 50%, and 75% percentiles from the bottom to the top.

as labels to establish a dataset for predicting MVNs connectivity.

The experimental results on the MVNs connectivity prediction dataset demonstrate that our proposed VC-Net, which employs DeepLabv3 as the backbone to extract MVNs morphological features, achieves the optimal performance, with an average Dice score of 0.81 and an IoU of 0.64. Furthermore, the experimental results demonstrate that the incorporation of the three independent modules of VQCL (the Projection module, the MLP module, and the Queue module), results in the optimal performance of VC-Net, with an average Dice score of 0.78, an IoU of 0.67, and an HD95 of 99.57. Finally, to address the severe imbalance in the MVNs connectivity dataset, where the background class accounts for an average area proportion of 62.7%, connectivity class 29.9%, and non-connectivity class 7.4%, we introduced the PPW method from our previous work. This resulted in the highest performance, with an average Dice score of 0.81, IoU of 0.70, and HD95 of 86.33. Consequently, our VC-Net employs DeepLabv3 as the backbone and incorporates the VQCL and PPW methods to enhance the performance of MVNs connectivity prediction.

Fluorescence imaging was performed on MVNs under both conventional culture and tumor microenvironment conditions. Parameters such as diameter and total length were measured and compared. In comparison to conventional culture conditions, MVNs under tumor microenvironment conditions exhibit a 22.85% reduction in average diameter (16 μm) and a 16.67% increase in average total length (8 mm). Furthermore, the predictions of VC-Net regarding the connectivity of MVNs under conditions of the tumor microenvironment demonstrate no significant discrepancy from the ground truth. The results indicate that, in comparison to conventional culture, the average number of connected class under tumor microenvironment conditions is reduced by 30.8%, while the number of non-connected class is increased by 37.3%. This indicates that the tumor microenvironment significantly alters MVNs connectivity, with some vascular channels narrowing and impeding fluid flow, resulting in a reduction in connected areas and an increase in non-connected areas.

## Experimental Section

### Fabrication of the Microfluidic Device

The microfluidic device was initially designed using computer-aided design (CAD) and printed to create a transparent mask with three gel channels and four medium channels (**Fig. 1**a, **Fig.**

S7). An 185 μm layer of SU-8 photoresist was coated onto a silicon wafer to create a mold with a positive relief pattern using soft lithography. Polydimethylsiloxane (PDMS, Sylgard 184, dowcorning) was mixed with an elastomer and curing agent at a 10:1 volume ratio, degassed, poured over the photolithography mold, and cured overnight at 60 °C. The cured PDMS was peeled from the mold, punched, and then subjected to plasma treatment (Harrick Plasma) in conjunction with a glass coverslip. The PDMS and glass were rapidly bonded together. Finally, the device was sterilized in an autoclave and then placed in an 80 °C drying oven overnight to render it hydrophobic.

**Materials and Cell Culture**

Human umbilical vein endothelial cells (HUVECs, MeiSen Cell) were passaged in endothelial growth medium (EGM-2, Lonza), and human lung fibroblasts (FBs, Procell) were passaged in fibroblast growth medium (FGM-2, Lonza). The cells were incubated at 37 °C with 5% $CO_2$ for a period of 2-3 days. Subsequently, the cells were dissociated using 0.25% trypsin/EDTA (Gibco) and seeded into the microfluidic device. Bovine fibrinogen (6 *mg/mL*, Sigma-Aldrich) and thrombin (4 *U/mL*, Sigma-Aldrich) were dissolved in EGM-2. The dissociated HUVECs and FBs were separately resuspended in the thrombin solution. Thereafter, the bovine fibrinogen and thrombin solutions were mixed at a 1:1 volume ratio. HUVECs (final concentration $6 \times 10^6$ *cells/mL*) and FBs (final concentration $1 \times 10^6$ *cells/mL*) were rapidly injected into the HUVECs-gel and FBs-gel channels filling ports, respectively. The device was placed in an incubator to polymerize for 15 minutes. Then, EGM-2 was introduced into the four medium channels and the culture was maintained for seven days, with medium channels every 12 hours. To prevent the diffusion of fluorescent substances into the hydrogel during perfusion, a suspension of HUVECs (final concentration $1 \times 10^6$ *cells/mL*) was introduced into both sides of the HUVECs channel prior to the formation of mature vessels. Subsequently, the device was inclined at a 90 ° angle for approximately 10 minutes, allowing gravity to facilitate the migration of the HUVECs to the medium-gel interface. The **Fig. S8** illustrates the simulation of medium in the hydrogel-filled chip, including speed and pressure. In the tumor microenvironment MVNs experiments, human liver cancer cells (HepG2s, Procell) were amplified in culture flasks using low-glucose medium (MEM, Servicebio) for one day. The supernatant was collected and stored as tumor cell medium (TCM) at -20 °C. During the experiment, the medium in the original chip was replaced with a mixture of EGM-2: TCM at a 1:1 ratio. Cell proliferation was characterized by adding CCK-8 (Dojindo) to the culture medium and measuring the absorbance using a microplate reader (BioTek).

**Immunocytochemistry and Imaging**

Samples of MVNs were first fixed with 4% paraformaldehyde (PFA, Beyotime) for 30 minutes, followed by permeabilization and blocking with 0.1% Triton X-100 (Sigma-Aldrich) + 10% goat serum (Beyotime) for 60 minutes to reduce nonspecific antibody binding. The samples were incubated at 4°C overnight with rabbit/IgG CD31 polyclonal antibody (1:100, Proteintech), followed by a 15-minute incubation with goat anti-rabbit IgG (H+L) cross-adsorbed secondary antibody (1:300, Thermo Fisher) to identify HUVECs. Following a 15-minute incubation period, the nuclei were stained with DAPI (Beyotime) for visualization purposes.

Confocal images were obtained using a confocal microscope (Nikon Eclipse Ti2 Spinning Disk) with 4x, 10x, 20x, and 40x objectives, and z-stack images were captured. An inverted phase-contrast fluorescence microscope (Nikon ECLIPSE Ts2-FL) was used to measure the parameters of vessels in the microfluidic device. ImageJ was employed for manual measurement of vessel diameter and fluorescence intensity of the fluorescent images, while Angiotool was used to analyze and quantify vessel area and length.

**MVNs Perfusion**

To confirm the connectivity of MVNs, culture medium was aspirated from one of the medium channels adjacent to the HUVECs-gel channel. Subsequently, 20 μL of 50 *μg/mL* 70 kDa FITC-dextran dye (Sigma-Aldrich) or 2 μm green polystyrene fluorescent microspheres (Yuanbiotech) was injected. Fluorescent perfusate was imaged using a confocal microscope, enabling the capture of images of bright-field and fluorescence in the same field of view. The fluorescent microsphere perfusion was recorded as a video at 26 frames per second using an inverted phase-contrast fluorescence microscope (Movie S1).

**Dataset Description and Implementation Details**

In the EGM-2 culture system, 31 independent MVNs culture and perfusion experiments were conducted, and corresponding bright field and fluorescence perfusion images were collected. Subsequently, the images will be subjected to preprocessing, including random flipping, cropping, discarding, and contrast normalization, with the objective of enhancing the data and obtaining a trainable dataset. This resulted in a total of 1984 images with a resolution of 1024 × 1024 pixels, comprising 992 bright field images and corresponding 992 fluorescence perfusion images. The dataset is partitioned into a training set comprising 704 pairs of images and a testing set comprising 288 pairs of images, with a ratio of 7:3. To provide supplementary data for TCM-related tasks, eight independent TCM-related experiments were conducted, yielding a total of 512 images after preprocessing (256 bright field images and 256 fluorescence perfusion images). Subsequently, the data were divided into a training set of 160 images and a testing set of 96 images, which were incorporated into the existing dataset. During the training phase, the SGD optimizer was employed for 1,000 epochs with a batch size of 8, an initial learning rate of 0.01, momentum of 0.9, and a weight decay of 1e-4.

To comprehensively and objectively evaluate the segmentation performance of the proposed method, we selected the following metrics for comparison:

$$Dice = \frac{2*|X \cap Y|}{|X|+|Y|} = \frac{2*TP}{2*TP+FP+FN} \quad (14)$$

$$IoU = \frac{TP}{TP+FP+FN} \quad (15)$$

$$HD95 = 0.95 * max(d_{XY}, d_{YX}) \quad (16)$$

$$ASD = \frac{1}{K}\sum_{i=0}^{K} D_i \quad (17)$$

$$Acc = \frac{TP + TN}{TP + TN + FP + FN} \quad (18)$$

where true positive (TP) and false positive (FP) are used to indicate the correct and incorrect identification of positives, respectively. Similarly, the terms true negative (TN) and false negative (FN) are used to indicate the correct and incorrect identification of negatives. X and Y represent the binary masks of the truth labels and the predicted results, respectively. The one-way Hausdorff distances from set X to set Y, denoted by $d_{XY}$, and from set Y to set X, denoted by $d_{YX}$. K represents the total number of pixels in the prediction results, and $D_i$ denotes the Euclidean distance from the *i*-th pixel in the prediction results to the nearest surface point in the truth labels. Dice and intersection over union (IoU) are sensitive to the internal fill of the prediction. Dice measures the overlap between the predicted results and the truth labels, while IoU measures the ratio of the intersection area to the union area of the predicted results and the truth labels. Both metrics are expressed on a scale from 0 to 1, with values closer to 1 indicating superior segmentation performance. The HD95 and Average Surface Distance (ASD) metrics are susceptible to the boundaries of the predicted results, reflecting the shape similarity between the predicted results and the truth labels. Both metrics range from 0 to infinity, with values closer to 0 indicating superior predicted performance. The accuracy metric (Acc) is employed to assess the overall classification accuracy of the model, with values ranging from 0 to 1. Values closer to 1 indicate superior performance.

**Statistical Analysis**

A statistical analysis of the values was conducted using analysis of variance and pairwise comparisons to determine if there were statistically significant differences among three or more data groups. The statistical tests were performed using Origin 2021b (OriginLab). The p-value thresholds were set as follows: *p < 0.05, **p < 0.01, and p ⩾ 0.05 (ns, not significant).

## Acknowledgements

This work was supported by the National Key R&D Program of China (No. 2022YFA1104800) and the National Natural Science Foundation of China (No. 52405327)

## Conflict of Interest

The authors declare no conflict of interest.